\documentclass[10pt,conference]{IEEEtran}

\IEEEoverridecommandlockouts
\usepackage{cite}
\usepackage{amsmath,amssymb,amsfonts}
\usepackage{algorithmic}
\usepackage{graphicx}
\usepackage{textcomp}
\usepackage{xcolor}
\usepackage{comment}
\usepackage{tcolorbox}

\usepackage{enumitem}
\usepackage{xurl}
\usepackage{blindtext}
\usepackage{listings}
\lstdefinelanguage{JavaScript}{
  keywords={typeof, new, true, false, catch, function, return, null, catch, switch, var, if, in, while, do, else, case, break},
  keywordstyle=\bfseries,
  ndkeywords={class, export, boolean, throw, implements, import, this},
  ndkeywordstyle=\color{darkgray}\bfseries,
  identifierstyle=\color{black},
  sensitive=false,
  comment=[l]{//},
  morecomment=[s]{/*}{*/},
  morestring=[b]',
  morestring=[b]"
}
\usepackage{multirow}

\newcommand{\ourapproach}[0]{AutoOAS}
\newcommand{\springdoc}[0]{springdoc-openapi}
\newcommand{\oas}[0]{OAS}

\newcommand{\projectCatwatch}[0]{CatWatch}
\newcommand{\projectCwa}[0]{CWA}
\newcommand{\projectOcvn}[0]{OCVN}
\newcommand{\projectOhsome}[0]{Ohsome}
\newcommand{\projectProxyPrint}[0]{ProxyPrint}
\newcommand{\projectQuartz}[0]{Quartz}
\newcommand{\projectUrCodebin}[0]{Ur-Codebin}

\def\BibTeX{{\rm B\kern-.05em{\sc i\kern-.025em b}\kern-.08em
    T\kern-.1667em\lower.7ex\hbox{E}\kern-.125emX}}
\begin{document}

\title{
Generating Accurate OpenAPI Descriptions from Java Source Code

\thanks{This work has been submitted to the IEEE for possible publication. Copyright may be transferred without notice, after which this version may no longer be accessible.}
}

\author{
	\IEEEauthorblockN{Alexander Lercher, Christian Macho, Clemens Bauer, Martin Pinzger}
	\IEEEauthorblockA{\textit{Department of Informatics Systems} \\
		\textit{University of Klagenfurt}\\
		Klagenfurt, Austria \\
		\{firstname.lastname\}@aau.at}
}

\maketitle

\begin{abstract}
Developers require accurate descriptions of REpresentational State Transfer (REST) Application Programming Interfaces (APIs) for a successful interaction between web services. The OpenAPI Specification (OAS) has become the de facto standard for documenting REST APIs. Manually creating an OpenAPI description is time-consuming and error-prone, and therefore several approaches were proposed to automatically generate them from bytecode or runtime information. 

In this paper, we first study three state-of-the-art approaches, Respector, Prophet, and \springdoc{}, and present and discuss their shortcomings. Next, we introduce \ourapproach{}, our approach addressing these shortcomings to generate accurate OpenAPI descriptions. It detects exposed REST endpoint paths, corresponding HTTP methods, HTTP response codes, and the data models of request parameters and responses directly from Java source code.

We evaluated \ourapproach{} on seven real-world Spring Boot projects and compared its performance with the three state-of-the-art approaches. 
Based on a manually created ground truth, \ourapproach{} achieved the highest precision and recall when identifying REST endpoint paths, HTTP methods, parameters, and responses. It outperformed the second-best approach, Respector, with a 39\% higher precision and 35\% higher recall when identifying parameters and a 29\% higher precision and 11\% higher recall when identifying responses. 
Furthermore, \ourapproach{} is the only approach that handles configuration profiles, and it provided the most accurate and detailed description of the data models that were used in the REST APIs.
\end{abstract}


\section{Introduction}
\label{sec:intro}

REpresentational State Transfer (REST)~\cite{fielding2000rest} Application Programming Interfaces (APIs) are widely used for communication between web services. 
Due to the loose coupling, service developers require documentation of the REST APIs to correctly implement API calls to other services.
The OpenAPI Specification (OAS)~\cite{openapisOpenAPISpecification} serves as the de-facto standard for describing REST APIs, and the resulting OpenAPI descriptions are typically shared with other development teams~\cite{LERCHER2024112110}.

The \textit{OpenAPI description} represents a REST API in two sections.
REST API endpoint paths and corresponding HTTP methods are located in the OAS section \textit{\#/paths}, and the data models received and returned by the REST API are contained in the section \textit{\#/components/schemas}. The \textit{\#} represents the OpenAPI description's root.
In the remainder of this paper, we use the term \textit{method} when referring to the unique combination of endpoint path and HTTP method and the term \textit{handler} when referring to the source code method handling the REST API call to clearly distinguish them.

Various tools~\cite{openapiOpenAPIToolsOpen} and research approaches~\cite{10.1145/3184558.3188740, PENG20181313, 10.1007/978-3-030-50578-3_40, 9159071, 10.1145/3293455, 10.1145/3639476.3639769} use OpenAPI descriptions, e.g., for visualization, %
testing, and test and client code generation, and rely on the description's accuracy. Manually creating OpenAPI descriptions is time-consuming and error-prone.
Many automated approaches~\cite{githubGitHubSpringdocspringdocopenapi, githubGitHubSwaggerapiswaggercore, 10.1109/ICSE48619.2023.00167, 10.1007/978-3-031-36889-9_4} require running the web service to generate the OpenAPI description. Hence, they require the domain knowledge to create a valid run configuration for the service and the infrastructure resources to run it.

Recently, two static analysis approaches emerged that generate REST API descriptions from Java source code and bytecode.
Huang et al.~\cite{10.1145/3597503.3639137} proposed Respector for generating OpenAPI descriptions from bytecode of web services created with Spring Boot~\cite{springSpringBoot} or Eclipse Jersey~\cite{eclipseee4jEclipseJersey}. 
They compared Respector to AppMap~\cite{appmapAppMap}, Swagger Core~\cite{githubGitHubSwaggerapiswaggercore}, SpringFox~\cite{githubGitHubSpringfoxspringfox}, and \springdoc{}~\cite{githubGitHubSpringdocspringdocopenapi} and outperformed all four state-of-the-art tools for generating methods, parameters, and responses. 
Cerny et al.~\cite{Cerny2024} proposed Prophet, an approach for statically analyzing REST APIs in Java Spring Boot projects. Prophet does not generate an OpenAPI description but creates a custom JSON output used as an intermediate format for visualizing microservice dependency graphs. %

However, in an evaluation of the existing static analysis approaches, Respector and Prophet,
we identified several limitations.
First, they do not consider Spring profiles~\cite{springProfilesSpring} and fail to correctly translate thrown exceptions to dedicated HTTP response status codes.
Moreover, they do not accurately report the data models of a REST API, which is a requirement for the various tools building on an OpenAPI description.

In this paper, we aim to understand and address the shortcomings of existing approaches and present \ourapproach{}, our static analysis approach for generating more accurate and detailed OpenAPI descriptions.
It detects exposed REST methods, parameters, responses, and data models directly from the Java source code. It considers Spring profiles and exception handling and accurately represents data models, including inheritance information.
We implemented \ourapproach{} for the Java framework Spring Boot. 
To the best of our knowledge, we are the first to present a static analysis approach for generating OpenAPI descriptions that considers Spring profiles and accurately describes the data models.
In this paper, we address the following research questions (RQs):
\newcommand{\rqOne}[0]{What are the shortcomings of state-of-the-art approaches for generating OpenAPI descriptions?} %
\newcommand{\rqTwo}[0]{How accurately does \ourapproach{} generate the OpenAPI description from Java Spring Boot source code compared to state-of-the-art approaches?}
\newcommand{\rqThree}[0]{What is the runtime performance of \ourapproach{} compared to state-of-the-art approaches?}

\begin{itemize}[leftmargin=1cm]
    \item[RQ1] \rqOne
    \item[RQ2] \rqTwo
    \item[RQ3] \rqThree
\end{itemize}

We compare \ourapproach{} to Respector~\cite{10.1145/3597503.3639137}, Prophet~\cite{Cerny2024}, and \springdoc{}~\cite{githubGitHubSpringdocspringdocopenapi} which are the state-of-the-art approaches for extracting the OpenAPI description from bytecode, source code, and runtime reflection, respectively. For the evaluation, we used an existing dataset containing seven Java Spring Boot projects~\cite{10.1145/3597503.3639137} and improved the ground truth of methods, parameters, and responses obtained from the source code.
The results show that \ourapproach{} obtained the highest precision and recall for identifying methods, parameters, and responses, and was the only approach that correctly described data model inheritance hierarchies.
In summary, this paper makes the following contributions: 
\begin{itemize}
    \item An improved dataset describing the REST APIs of seven Java Spring Boot projects.
    \item An analysis of the shortcomings of the state-of-the-art approaches for generating OpenAPI descriptions.
    \item Our approach \ourapproach{} addressing the shortcomings and thereby generating more accurate OpenAPI descriptions.
\end{itemize}

The remainder of this paper is structured as follows.
Section~\ref{sec:gt-improvements} describes the dataset used for the evaluations and the improvements we performed.
Section~\ref{sec:shortcomings} discusses the shortcomings of existing approaches for generating OpenAPI descriptions.
Our approach \ourapproach{}, which addresses the shortcomings, is presented in Section~\ref{sec:approach}.
In Section~\ref{sec:eval-oas}, we evaluate the precision, recall, and data model representation quality of the OpenAPI descriptions generated by \ourapproach{} compared to existing approaches. 
In Section~\ref{sec:eval-runtime}, we evaluate its runtime performance. 
Section \ref{sec:discussion} discusses the implications and threats to validity, and Section \ref{sec:relatedwork} presents related work. Finally, Section~\ref{sec:conclusion} concludes the paper.

\section{Evaluating and Improving the Ground Truth}
\label{sec:gt-improvements}

For our evaluation of existing approaches, we used the dataset curated by Huang et al.~\cite{10.1145/3597503.3639137} containing seven Java Spring Boot projects and a document reporting their ground truth (GT), i.e., the projects' methods, parameters, and responses. 
Table \ref{tab:descriptive-statistics} lists these projects and provides descriptive statistics.

\begin{table}[htbp]
\caption{Descriptive statistics of the seven Spring Boot projects.}
\begin{center}
\begin{tabular}{|l|r|r|r|r|}

\hline
Project   & Java LoC  & \#methods & \#parameters & \#responses \\
\hline

\projectCatwatch{}~\cite{githubEMBjdk_8_mavencsrestoriginalcatwatchMaster}      & 6,454    & 14      & 36         & 19     \\
\projectCwa{}~\cite{githubEMBjdk_11_mavenemembeddedrestcwaverificationMaster}           & 3,616    & 6       & 16         & 16      \\
\projectOcvn{}~\cite{githubEMBjdk_8_mavencsrestguiocvnMaster}          & 28,099   & 278     & 5,002      & 278   \\
\projectOhsome{}~\cite{githubGitHubGIScienceohsomeapi}        & 10,597   & 159     & 1,937      & 159       \\
\projectProxyPrint{}~\cite{githubEMBjdk_8_mavencsrestoriginalproxyprintMaster}    & 6,052    & 75      & 154        & 101       \\
\projectQuartz{}~\cite{githubGitHubFabioformosaquartzmanager}        & 3,883    & 14      & 15         & 20     \\
\projectUrCodebin{}~\cite{githubGitHubMathewEstafanousUrCodebinAPI}     & 1,962    & 6       & 14         & 12   \\

\hline

\end{tabular}
\label{tab:descriptive-statistics}
\end{center}
\vspace{-0.6cm}
\end{table}

During our evaluation of existing approaches, we found multiple errors in the GT %
by manually comparing the OpenAPI descriptions generated by existing approaches with the projects' source code. All detected errors were discussed by at least two authors to avoid bias. 
In the following, we describe the errors and the improvements that we performed to create our improved GT+ dataset.

\subsection{Profiles and Methods}
\label{sec:gt-corrections:paths}
Spring Boot allows developers to segregate parts of the web service and its REST API at runtime by using Spring Profiles~\cite{springProfilesSpring}. In particular, it provides an \texttt{@Profile} class annotation assigning the corresponding class to one or multiple profiles. At runtime, it only instantiates the classes of active profiles based on the service configuration.
For instance, the \projectCwa{} project consists of two Spring profiles called \textit{external} and \textit{internal}. 
Both profiles contain the same method and due to Spring Boot's constraint of unique methods, the two profiles cannot be simultaneously active at runtime. GT missed this constraint. We improved it by creating two separate service descriptions to consider this runtime behavior in our GT+. 
Moreover, GT contained incorrect paths for project \projectUrCodebin{}. In particular, we found that \textit{/api} prefixes were missing for all endpoint paths in the GT. We fixed the paths in GT+.

\subsection{Parameters}
\label{sec:gt-corrections:parameters}
We noticed that the GT does not contain parameter locations, e.g., path or query. Furthermore, it represents parameter objects as individual parameters based on the object's fields. For instance, it represents a single parameter object with three fields as three individual parameters. 
As a single exception, the GT reports one object parameter in project \projectCwa{} with the object's variable name instead of its only field's name. 
We renamed the parameter to its field name in the GT+ and adopted the flat representation of the GT. This enables us to evaluate the completeness of the object parameter fields generated by the approaches.

In project \projectCatwatch{}, we discovered that the GT reported a parameter name as \texttt{sort\_by} instead of \texttt{sortBy} and we updated our GT+ accordingly.  
Furthermore, one request body parameter for project \projectCatwatch{} is missing in the GT. %
Similarly, project \projectQuartz{} contains multiple request body parameters missed by the GT. 
After verifying their appearances in the source code, we added all of them to the GT+.

\subsection{Responses}
\label{sec:corrected-gt:responses}
We performed multiple improvements on the responses for two reasons. %

\subsubsection{Default response code}
\label{sec:gt-corrections:parameters:response-codes}
The authors of the GT assumed that methods with the void return type and methods returning null translate to a \textit{204 No Content} response code. However, Spring Boot returns \textit{200 OK} by default and only returns \textit{204} if the status code is explicitly set. We found and fixed this error in the projects \projectOcvn{}, \projectOhsome{}, and \projectProxyPrint{}.

\subsubsection{Exception handling}
Spring Boot provides \texttt{@ExceptionHandler} annotations to translate uncaught Java exceptions in handlers to HTTP response codes, typically defined in a second annotation \texttt{@ResponseStatus}. However, the GT does not report many translated error response codes. 
For instance, in project \projectCatwatch{}, four methods always throw an \textit{UnsupportedOperationException}. %
The exception translates to the HTTP status code \textit{403 Forbidden}, which we added to the GT+. We fixed such errors also in the projects \projectProxyPrint{}, \projectQuartz{}, and \projectUrCodebin{}.

\section{Shortcomings of existing approaches}
\label{sec:shortcomings}

In this section, we evaluate three state-of-the-art approaches for generating OpenAPI descriptions, namely Respector~\cite{10.1145/3597503.3639137}, Prophet~\cite{Cerny2024}, and \springdoc{}~\cite{githubGitHubSpringdocspringdocopenapi}, to answer RQ1. 

\subsection{Evaluation setup}
\label{sec:shortcomings:setup}
We generated the OpenAPI descriptions for each project with each approach and compared them to GT+.
Furthermore, we manually evaluated the data model representation quality based on the correct translation of Java types into OAS types and the correctness of the model schemas compared to the Java classes.

For Repector, we compiled the source code of each project to obtain the byte code necessary for its analysis. We used the compilation information contained in the GT dataset to ensure that we evaluated the same state as in the original evaluation of Respector.
For \springdoc{}, we took the OpenAPI descriptions that were generated by Huang et al.~\cite{10.1145/3597503.3639137}. As in Huang et al., we could not find a valid run configuration for the three projects \projectOcvn{}, \projectProxyPrint{}, and \projectQuartz{} and, hence, could not generate the OpenAPI descriptions for them.
Prophet generates custom JSON files as analysis output. It does not report the actual parameter and response schemas but only the string literals in the Java method signatures. A method description for project \projectCatwatch{} is shown in Listing~\ref{lst:prophet-output}.

For our evaluation, we converted Prophet's custom JSON files into OpenAPI descriptions containing the original endpoint paths (Line~4 in Listing~\ref{lst:prophet-output}), HTTP methods (Line~1), parameters (Line~2), and success responses (Line~3). %
Additionally, we analyzed each parameter to identify its Java annotations, type, and name. We analyzed the annotations to output request body parameters in the corresponding OpenAPI section. We output Prophet's \texttt{returnType} property as the response schema for response code \textit{200 OK}. We selected Spring Boot's default response code because Prophet itself does not identify any response codes.
Note, the enhanced output contains no schema information but only the class names extracted from Prophet's string literals. In summary, we not only correctly converted the extracted information of Prophet but even enhanced its output by considering the \texttt{@RequestBody} annotation and setting the default response code for successful responses. 
We provide the conversion script for Prophet in our replication package~\cite{replication-package}.

\begin{figure}[t]
\begin{lstlisting}[caption={
A snippet of Prophet's output describing one method of \projectCatwatch{}.
}, label={lst:prophet-output}, language=JavaScript]
  { "httpMethod": "GET",
     "arguments": "[@RequestBody(required = false) String scoringProject, @RequestHeader(value = \"X-Organizations\", required = false) String organizations]",
     "returnType": "java.lang.String",
     "path": "/config/scoring.project",
  [...] }
\end{lstlisting}
\vspace{-.6cm}
\end{figure}

In the following, we describe the common reasons for incorrect identifications of methods, parameters, and responses compared to the GT+ and incorrect data model representations compared to the source code.

\subsection{Methods}
Existing approaches did not always extract methods correctly 
because of the following four reasons.

\subsubsection{Spring profiles}
\label{sec:qual:spring-profiles}
Project \projectCwa{} uses Spring profiles to distinguish its runtime behavior (cf. Section~\ref{sec:gt-corrections:paths}). The two profiles named \textit{external} and \textit{internal} each contain a handler exposing the same method
but different parameters, response codes, and even a different response body. 

Respector does not support Spring Boot profiles and only generated a single OAS description containing one of the two duplicate methods and all other methods of both profiles.
Prophet correctly detected the duplicate method but reported all methods in the same file without information about their accessibility at runtime. Notably, the \oas{} prohibits duplicate methods for OpenAPI descriptions, and Prophet could only report both methods in the same file because of its custom output format.

\springdoc{} did not report any methods or data models in the OpenAPI description of project \projectCwa{} because we could not run the project with any profile. 
We argue that \springdoc{} reports the correct subset of methods for the profile active at runtime because only these handlers are loaded at startup. However, therefore \springdoc{}'s analysis requires re-starting the service for each profile.

\subsubsection{Constants in paths}
\label{sec:shortcomings:constants}
Prophet could not resolve constants in endpoint paths but instead reported the string literal as the path, e.g., \textit{/api/ENDPOINT\_NAME}. This especially affected the projects \projectCwa{} and \projectQuartz{}, which extensively use constants.

\subsubsection{Regular expression constraints}
Project \projectOcvn{} uses regular expressions for several path parameters, encoded in the path string. 
Respector correctly generated the path but ignored the constraint in the parameter description. 
Similar to handling constants, Prophet wrongly retained the literal path string as the path.

\subsubsection{Request mapping annotations}
Prophet incorrectly identified the methods of handlers annotated with \texttt{@RequestMapping} in multiple projects.
Sometimes, it identified only one method when the annotation defined multiple methods.
It also misclassified some HTTP methods as \textit{GET} when the annotation defined another one.
Furthermore, Prophet missed all \texttt{@RequestMapping} annotations which did not set any HTTP method, e.g., in project \projectProxyPrint{}. In this case, Spring Boot exposes the endpoint path with all HTTP methods, which was correctly reported by Respector.

\subsection{Parameters}
\label{sec:misclassification:endpoint-parameters}
We identified the following three reasons for misidentifying parameters.

\subsubsection{Model attribute annotations}
Project \projectOcvn{} extensively uses the \texttt{@ModelAttribute} annotation, which maps individual parameters to a single Java object.
After inspecting Respector's OpenAPI description, we discovered that it reported such parameters as a single request body object. However, Spring Boot does not bind the request body of the HTTP request to a model attribute object when receiving an API call~\cite{springModelAttributeSpring}. Instead, it expects one path, query, or form parameter per object field, and builds the Java object correspondingly. Hence, API calls according to Respector's OpenAPI description fail. 
We could not execute %
\springdoc{} for project \projectOcvn{} to evaluate its description of the \texttt{@ModelAttribute} annotation, and Prophet wrongly reported the model attribute parameter as a single parameter.

\subsubsection{Json property annotations}
Respector and Prophet both missed the \texttt{@JsonProperty} annotations in project \projectUrCodebin{}, which translate Java class field names to data model field names. 
Only \springdoc{} handled the annotation correctly and translated all parameter names.

\subsubsection{Request body annotations}
Respector missed one \texttt{@RequestBody} annotation in project \projectCatwatch{} and four in project \projectProxyPrint{}. 
We could not explain why Respector missed these five instances because we could not observe any differences from other request body annotation occurrences.
Prophet also missed the same annotation in project \projectCatwatch{}, whereas it detected the four annotations in project \projectProxyPrint{}. %
Only \springdoc{} correctly reported the parameter in project \projectCatwatch{}, but it could not generate an OpenAPI description for project \projectProxyPrint{}.

\subsection{Responses} 
We identified the following four reasons for misidentifying responses.

\subsubsection{Exception handler annotations}
All three approaches did not fully consider the \texttt{@ExceptionHandler} and \texttt{@ResponseStatus} annotations to translate exceptions to response codes (cf. Section~\ref{sec:corrected-gt:responses}).
For instance, the four methods in project \projectCatwatch{} throwing the \texttt{Unsupported} \texttt{OperationException} were incorrectly identified by all three approaches. While the exception translates to \textit{403 Forbidden}, Respector reported it as \textit{500 Internal Server Error}, and Prophet and \springdoc{} reported \textit{200 OK}.

\subsubsection{Propagated exceptions}
In the projects \projectCatwatch{} and \projectQuartz{}, several handlers propagate thrown runtime exceptions from their callees. 
Respector detected the exceptions but could not translate them to the correct response codes based on the \texttt{@ExceptionHandler} annotations.
Prophet could not identify the runtime exceptions because they are not part of the handler's signature, and \springdoc{} could not identify them from runtime reflection.

\subsubsection{Manual exception handling} 
We discovered that project \projectUrCodebin{} contains custom logic to translate thrown exceptions to response codes in the \texttt{@ExceptionHandler}-annotated methods.
Respector reported \textit{500 Internal Server Error} instead of the manually translated response codes. %
Prophet and \springdoc{} did not detect any exceptions and consequently did not report any error response codes.

\subsubsection{Wrong success response codes} %

Respector misidentified the success response codes of void handlers and handlers returning null as \textit{204 No Content} (cf. Section~\ref{sec:gt-corrections:parameters:response-codes}). 
Prophet did not describe any response codes. We manually set the response codes for Prophet to \textit{200 OK} in our conversion script. %
\springdoc{} did not suffer from this shortcoming.

\subsection{Data model representation}

We reiterate that the GT+ represents all parameter object fields as individual parameters to evaluate the completeness of the objects generated by the approaches (cf. Section~\ref{sec:gt-corrections:parameters}).
Furthermore, 
the GT+ only contains the response type, e.g., object, array, or primitive type, but not the schemas of response objects. 
Hence, as part of this evaluation, we explicitly assess the representation quality of data models in the generated OpenAPI descriptions. We define describing the data models in the \textit{\#/components/schemas} section and retaining their inheritance information as quality attributes.
We did not consider Prophet for most of this comparison because it generates custom JSON structures instead of OpenAPI descriptions and does not analyze the schemas of data models.

\subsubsection{Objects and inheritance}
\label{sec:shortcomings:flattened-objects}
In addition to regular objects, 
the projects \projectProxyPrint{} and \projectQuartz{} use derived objects, i.e., objects inheriting fields from a superclass, as request body parameters, and the projects \projectCwa{}, \projectProxyPrint{}, and \projectQuartz{} return derived objects as responses.

Respector generated anonymous schemas containing all object fields directly in the corresponding parameter or response section of the OpenAPI description. It did not generate any explicit schema information in the \textit{\#/components/schemas} section and thereby discarded the semantic information and relationships of the data models.
Furthermore, it flattened all fields of derived objects used as parameters and thereby discarded the inheritance information.
We also found that Respector did not consider the object hierarchy of responses at all. For instance, a method in project \projectCwa{} returns a data model named \texttt{InternalTestResult} with multiple inherited fields. However, Respector only reported a single field, which is declared directly in the class, without considering the inherited fields.

\springdoc{} placed data model objects in the \textit{\#/components/schemas} section of the OpenAPI description and referenced them in the corresponding parameter and response sections. 
However, we could not generate any OpenAPI descriptions containing derived objects from the dataset's projects. To evaluate the representation quality of \springdoc{}, we created a minimal test project containing derived objects.
We found that it flattened all inherited fields of the object into a single schema and, similar to Respector, discarded all inheritance information.

\subsubsection{Map data structure}
The projects \projectCatwatch{}, \projectOcvn{}, and \projectProxyPrint{} use the map data structure, also known as dictionary, in parameters and responses.  
The \oas{} describes maps as objects. It implicitly defines the map's key type as a string and provides the \texttt{additionalProperties} keyword for the map's value type~\cite{openapisOpenAPISpecificationMapDict}.  
We found that Respector did not describe the value type but only reported the data structure as a regular object without any other information.
In contrast, \springdoc{} correctly reported the type of the values with the \texttt{additionalProperties} keyword.

\noindent{\bf Answer RQ1}:
We identified several shortcomings with existing approaches that impact the quality of OpenAPI descriptions extracted from Spring Boot projects. They range from ignoring spring profiles, over ignoring or incorrectly handling several Spring Boot annotations and exception handling, to incorrectly representing data schemas. This hampers the application of existing approaches to Spring Boot projects and motivates a new, improved approach that we introduce in the next section.

\section{\ourapproach{} approach}
\label{sec:approach}
Based on the identified shortcomings, %
we propose our approach named \ourapproach{} for generating accurate OpenAPI descriptions from Java source code.
\ourapproach{} uses Spoon~\cite{pawlak:hal-01169705} to parse and statically analyze the Java source code, its inheritance hierarchy, and annotations. It does not require byte code or a running, accessible service, which simplifies the OpenAPI generation process by omitting compilation, execution, and runtime dependency management.
Currently, our approach is limited to Spring Boot projects. However, it can be extended to support other vendors and frameworks.

\ourapproach{} identifies all Java classes containing REST method definitions and groups them based on their assignment to Spring profiles. It generates one OpenAPI description for each profile. For this, it identifies the methods, parameters, and responses from the source code definitions and uses the information to generate the OAS \textit{\#/paths} section. Then, it detects the data models referenced as parameter and response objects and generates the corresponding schema information in the OAS \textit{\#/components/schemas} section.
Notably, if a service configuration defines multiple active profiles at once, merging the individual OpenAPI descriptions is trivial because all methods must be unique at runtime, and data model classes must be unique at compile time. 
Therefore, OpenAPI descriptions are mergeable with set unions of their \textit{\#/paths} sections, which are disjunct, and \textit{\#/components/schemas/} sections, which are either disjunct or contain identical duplicates.

In the following, we present the three stages of \ourapproach{} in detail, maintaining the distinction between \textit{method} and \textit{handler} (cf. Section~\ref{sec:intro}).

\subsection{Source code parsing}
\label{meth:source-code-filtering}
\ourapproach{} parses the project's source code with Spoon to identify all Java classes that contain method definitions.
For this, \ourapproach{} detects annotations marking the parsed Java classes that expose methods as \textit{controllers}, such as \texttt{@RestController}. 
It assigns each controller class to their assigned Spring profile set(s) or to all profiles, including the default profile, if they are not explicitly assigned to at least one.
\ourapproach{} then generates one individual OpenAPI description for each profile as described in the following sections.

\subsection{Generating OpenAPI methods, parameters, and responses}
In this stage, \ourapproach{} generates the \textit{\#/paths} section of the OpenAPI description by analyzing all method definitions in the controller classes of a single Spring profile.

\subsubsection{Methods}
\ourapproach{} considers superclasses of controllers because they potentially contain additional configurations or methods. 
Starting from the controller classes, \ourapproach{} traverses each controller's inheritance hierarchy and identifies all Java methods, i.e., handlers, annotated with \texttt{@RequestMapping} or specialized HTTP method mapping annotations, e.g., \texttt{@GetMapping}, as method definitions. 
\ourapproach{} extracts the endpoint path and HTTP methods directly from the handler's mapping annotations. 

\subsubsection{Parameters}
Next, \ourapproach{} analyzes the parameters. %
It identifies each parameter name and schema, i.e., Java type information, from the handler's signature. %
If a parameter annotation explicitly defines the parameter name, e.g., \texttt{@RequestParam("param\_name")}, \ourapproach{} renames it accordingly. It also resolves the parameter location, i.e., path, query, or header, from the annotation. 
If a path parameter contains a regular expression (regex) constraint following the syntax \texttt{/\{parameter:regex\}}, \ourapproach{} removes the regex from the path and adds it to the \texttt{pattern} constraint field of the parameter description.

\ourapproach{} converts parameter objects annotated with \texttt{@ModelAttribute} into individual query parameters based on the object fields. It also considers inherited fields by recursively iterating the object's inheritance hierarchy.
The \texttt{@RequestBody} annotation marks a special parameter transferred in the HTTP request body, and \ourapproach{} uses the dedicated OAS keyword \texttt{requestBody} to describe it. 
Notably, \ourapproach{} does not describe object parameter schemas in-line but generates a reference to a named schema in the \textit{\#/components/schemas/} section of the OpenAPI description by using the \texttt{\$ref} keyword. This enables the reusability of the service's data models.

\subsubsection{Responses}
\ourapproach{} analyzes each handler's body for \texttt{return} and \texttt{throws} statements to identify the responses.
Handlers may return regular Java objects or \texttt{ResponseEntity<T>} objects, which wrap Java objects of type \texttt{T} and potentially set HTTP response codes.
\ourapproach{} detects response codes returned by \texttt{ResponseEntity} objects from the source code. 
In case the handler has a void return type or no particular \texttt{ResponseEntity} or explicit \texttt{@ResponseStatus} annotation could be identified, \ourapproach{} sets the response code to \textit{200 OK}.
\ourapproach{} extracts the data model schema of successful responses from the handler's return type and references its named schema in the \textit{\#/components/schemas/} section.

Handlers may throw exceptions with the \texttt{throws} statement, resulting in HTTP client and server error codes.
Spring Boot provides \texttt{@ExceptionHandler} and \texttt{@ResponseStatus} annotations to translate Java exceptions to HTTP response codes (cf. Section~\ref{sec:corrected-gt:responses}).
The Java method containing these annotations is either declared inside a controller to translate the exceptions of all handlers inside the same controller, i.e., local exception handling, or declared in a separate class annotated with \texttt{@ControllerAdvice} for translating the exceptions of any handler, i.e., global exception handling. 

\ourapproach{} detects such exception-handling behavior.
For this, it detects all classes annotated with \texttt{@ControllerAdvice} during the parsing stage described in Section~\ref{meth:source-code-filtering} to store the global exception handlers.
When encountering a \texttt{throws} statement in a handler's body, \ourapproach{} translates the thrown exception to a response code, with precedence on local exception handling. %
It first tries to identify a locally declared exception handler Java method for the thrown exception and resolves the corresponding response code. If it could not translate the exception, it searches the global exception handler Java methods for a matching exception to resolve the response code.
Unresolved exceptions result in an \textit{500 Internal Server Error} and are reported accordingly.

\subsection{Generating OpenAPI data models}
\label{meth:schemas}
In this stage, \ourapproach{} 
generates the data models, i.e., the data objects transferred during API calls.
We differentiate between two types of data models: simple types and named schemas.
\ourapproach{} puts simple types directly into the corresponding parameter or response description and named schemas into the \textit{\#/components/schemas/} section of the OpenAPI description~\cite{openapisOpenAPISpecificationComponentsObject}.

\subsubsection{Simple types}
\ourapproach{} supports Java's built-in primitive and reference types, e.g., int or String. 
It maps primitive types and their wrapper objects to the basic types of the OAS~\cite{openapisOpenAPISpecificationDataTypes}, e.g., int to \textit{integer}, double to \textit{number}. 
It converts arrays and lists to the OAS \textit{array} type, and explicitly lists Java enum constants in the OpenAPI description. 
It reports the map data structure, also known as the dictionary data structure, as an \textit{object} and uses the \texttt{additionalProperties} keyword to describe the value type.

\subsubsection{Named schemas}
\label{meth:schemas:named-schemas}
\ourapproach{} reports custom Java classes referenced in parameters and responses as named schemas in the OpenAPI description \textit{\#/components/schemas/} section.
For this, \ourapproach{} records all Java classes referenced during the previous stage, which are the classes required to correctly interact with the described methods.
\ourapproach{} generates the named schemas by describing the fields of each data model class. Additionally, it marks required fields of the schema with the OAS \texttt{required} keyword. \ourapproach{} considers fields as required if they contain the \texttt{@NotNull} or \texttt{@NotEmpty} validation annotations or are primitive Java types which are not nullable, e.g., int.

If the data model class is a derived class, i.e., a class inheriting fields from a superclass, \ourapproach{} uses the \texttt{allOf} keyword~\cite{openapisOpenAPISpecificationPolymorphism} of the OAS to create a combined schema which contains the schema of the current class and a reference to its superclass. 
It then generates the named schema of the superclass and, if necessary, references its superclass again. Thereby, \ourapproach{} recursively scans the inheritance tree and retains the transitive inheritance information in the OpenAPI description instead of creating flat data model schemas, i.e., describing all fields in the same schema.

We argue that explicitly referenced and inheritance-aware schemas are necessary for correctly understanding and interacting with a service. 
First, referencing named schemas in parameters and responses helps developers understand the relationships between calls. 
For instance, two methods using the same schemas potentially belong to the same workflow or are potential API call chain candidates. 
Second, code generation tools require named schemas to correctly generate API consumers. 
For instance, they would generate multiple equivalent classes from duplicate anonymous schemas that are not compatible at runtime and, hence, not reusable in API call chains.

\subsubsection{Response wrappers}
Finally, \ourapproach{} automatically handles Spring's \texttt{ResponseEntity<T>} and \texttt{Deferred} \texttt{Result<T>} response wrappers by extracting and reporting the type parameter \texttt{T}. If the type parameter is not specified, \ourapproach{} uses a special named schema called \textit{UNSPECIFIED\_TYPE}, and if it cannot infer the type, e.g., because the defining class is located in an inaccessible source package, \ourapproach{} creates a named schema from the class name and reports the package name in the OAS's \texttt{externalDocs} field.

\section{Evaluation of Precision, Recall, and \\ Data Model Quality}
\label{sec:eval-oas}

In this section, we answer RQ2 by evaluating \ourapproach{} on the GT+ described in Section~\ref{sec:gt-improvements}.
We compare the precision and recall of \ourapproach{}, Respector, Prophet, and \springdoc{} for identifying methods, parameters, and responses. Furthermore, we evaluate the data model representation quality, i.e., correctly referencing data models in the \textit{\#/components/schemas} section and retaining inheritance information. To generate the OpenAPI descriptions of Respector, Prophet, and \springdoc{}, we reused the setup from Section~\ref{sec:shortcomings:setup}.

We created a script to automatically calculate precision and recall. 
First, for each approach, it transforms the generated OpenAPI descriptions into lists of methods, parameters, and responses per project. Next, it compares the list entries to the methods, parameters, and responses in the GT+. An identified method is a true positive (TP) if the endpoint path and HTTP method combination exists in the GT+, an identified parameter is a TP if the method and parameter name match with the GT+, and an identified response is a TP if the method and response code match.
The script also counts false positives (FP), i.e., identified methods, parameters, or responses not existing in the GT+, and false negatives (FN), i.e., not identified methods, parameters, or responses that do exist in the GT+. 
With these results, it calculates precision and recall as follows:
\[
\text{Precision} = \frac{TP}{TP + FP} \text{, }
\text{Recall} = \frac{TP}{TP + FN}.
\]
We define the precision without any TP or FP findings as 0\% to penalize approaches not reporting any methods, parameters, or responses.
The results for each project for each approach are presented in Table~\ref{tab:results-corrected-gt}.
In the following, we describe them in detail. 

\newcommand{\hl}[1]{\textbf{#1}}
\newcommand{\zero}[1]{{\color{gray}#1}}

\begin{table*}[htbp]
\caption{Results of the four approaches with GT+. The GT+ column contains the number (\#) of methods, parameters, and responses for each project.
Unsuccessful analyses are marked with an "x". The best results per project are highlighted in \hl{bold}.}

\begin{center}

\begin{tabular}{|ll|r|rr|rr|rr|rr|}
\hline
                         \multicolumn{2}{|c|}{Project}   &  \multicolumn{1}{|c|}{GT+}        & \multicolumn{2}{|c|}{\ourapproach{}}                                   & \multicolumn{2}{|c|}{Respector}                              & \multicolumn{2}{|c|}{Prophet}                                 & \multicolumn{2}{|c|}{springdoc-openapi}  \\
                            &          & \multicolumn{1}{c|}{\#}  & \multicolumn{1}{c}{Precision} & \multicolumn{1}{c|}{Recall} & \multicolumn{1}{c}{Precision} & \multicolumn{1}{c|}{Recall} & \multicolumn{1}{c}{Precision} & \multicolumn{1}{c|}{Recall} & \multicolumn{1}{c}{Precision} & \multicolumn{1}{c|}{Recall}  \\

\hline
\multirow{3}{*}{\projectCatwatch{}}     & methods       & 14    & \hl{1.00}                          & \hl{1.00}                       & \hl{1.00}                          &  \hl{1.00}                       & 0.60                          & 0.21                       &  1.00                        & 0.57                                     \\
                              & parameters  & {36} &   \hl{1.00}                          &   \hl{1.00}                       &   1.00                          & {0.97}                       & \zero{0.00}                          & \zero{0.00}                       &   1.00                      & {0.33}                                           \\
                              & responses       & {19}     &   \hl{1.00}                          &  \hl{0.79}                      & {0.79}                          &  0.79                       & 0.20                          & {0.05}                       & {0.62}                          & {0.42}                                  \\
\hline
\multirow{3}{*}{\projectCwa{} external} & methods     & {3}         &   \hl{1.00}                          &   \hl{1.00}                       & {0.60}                          &  1.00                      & \zero{0.00}                          & \zero{0.00}                       & \zero{0.00}                            & \zero{0.00}                                   \\
                              & parameters   & {10}      &   \hl{1.00}                          &   \hl{1.00}                       & {0.69}                          & {0.90}                       & \zero{0.00}                          & \zero{0.00}                       & \zero{0.00}                            & \zero{0.00}                                     \\
                              & responses         & {8}   &  \hl{0.86}                          &  \hl{0.75}                       & {0.22}                          & {0.25}                       & \zero{0.00}                          & \zero{0.00}                       & \zero{0.00}                            & \zero{0.00}                                    \\
\hline
\multirow{3}{*}{\projectCwa{} internal} & methods    & {3}   &   \hl{1.00}                          &   \hl{1.00}                       & {0.60}                          &  {1.00}                       & \zero{0.00}                          & \zero{0.00}                       & \zero{0.00}                             & \zero{0.00}                                           \\
                              & parameters   & {6}  &   \hl{1.00}                          &   \hl{1.00}                       & {0.38}                          & {0.83}                       & \zero{0.00}                          & \zero{0.00}                       & \zero{0.00}                            & \zero{0.00}                                         \\
                              & responses    & {8}     & \hl{0.88}                          & \hl{0.88}                       & {0.33}                          & {0.38}                       & \zero{0.00}                          & \zero{0.00}                       & \zero{0.00}                            & \zero{0.00}                                       \\
\hline
\multirow{3}{*}{\projectOcvn{}}         & methods    & 278        &  \hl{1.00}                          &  \hl{1.00}                       &  \hl{1.00}                          &  \hl{1.00}                       & 0.95                          & 0.45                       & x                             & x                                       \\
                              & parameters       & 5,002       &  \hl{1.00}                          & \hl{0.95}                       & {0.06}                          & {0.06}                       & 0.06                          & \zero{0.00}                       & x                             & x                               \\
                              & responses       & 278  &   \hl{1.00}                          &   \hl{1.00}                       & {0.81}                          & {0.89}                       & {0.95}                          & {0.45}                      & x                             & x                                        \\
\hline
\multirow{3}{*}{\projectOhsome{}}       & methods        & 159         &  \hl{1.00}                          &  \hl{1.00}                       &  \hl{1.00}                          &  \hl{1.00}                       &  {1.00}                          & 0.50                       & 0.88                          &  {1.00}                               \\
                              & parameters       & 1,937    & \zero{0.00}                            & \zero{0.00}                       &  \hl{1.00}                          & \hl{0.99}                       & \zero{0.00}                          & \zero{0.00}                       & \zero{0.00}                            & \zero{0.00}                               \\
                              & responses    & {159}      &   \hl{1.00}                          &   \hl{1.00}                       & {0.35}                          & {0.85}                       &   {1.00}                          & {0.50}                       & {0.88}                          &   {1.00}                                   \\
\hline
\multirow{3}{*}{\projectProxyPrint{}}   & methods      & 75     &  \hl{1.00}                          &  \hl{1.00}                       &  \hl{1.00}                          &  \hl{1.00}                       & 0.99                          & 0.89                       & x                             & x                                        \\
                              & parameters    & {154}    &   {1.00}                          & {0.80}                       &  \hl{1.00}                          & \hl{0.95}                      & {0.31}                          & {0.22}                       & x                             & x                                       \\
                              & responses    & {101}       &   \hl{1.00}                          &   \hl{1.00}                       & {0.97}                          & {0.97}                      & {0.99}                          & {0.66}                       & x                             & x                                     \\
\hline
\multirow{3}{*}{\projectQuartz{}}       & methods        & 14     &  \hl{1.00}                          &  \hl{1.00}                       &  \hl{1.00}                          &  \hl{1.00}                       & 0.07                          & 0.07                       & x                             & x                                      \\
                              & parameters     & {15}  &   \hl{1.00}                          &  \hl{1.00}                       &   \hl{1.00}                          &  \hl{1.00}                       & \zero{0.00}                          & \zero{0.00}                       & x                             & x                                          \\
                              & responses       & {20}  &   {1.00}                          & {0.75}                       &   \hl{1.00}                          &  \hl{0.80}                       & 0.07                          & {0.05}                       & x                             & x                                         \\
\hline
\multirow{3}{*}{\projectUrCodebin{}}   & methods    & 6           &  \hl{1.00}                          &  \hl{1.00}                       &  \hl{1.00}                          &  \hl{1.00}                       &  \hl{1.00}                          &  \hl{1.00}                       &  \hl{1.00}                          &  \hl{1.00}                                  \\
                              & parameters      & 14   & 0.43                          & 0.43                       & 0.43                          & 0.43                       & 0.13                          & 0.07                       &  \hl{1.00}                          &  \hl{1.00}                                    \\
                              & responses      & {12}  & {0.55}                          & {0.50}                       & {0.45}                          & {0.42}                       &   \hl{1.00}                          &  \hl{0.50}                        &   \hl{1.00}                          & \hl{0.50}                            \\
\hline
\hline
\multirow{3}{*}{Overall} & methods    & 552                           &  \hl{1.00}                       &  \hl{1.00}                          &  \hl{1.00}                       &  \hl{1.00}                          & 0.93                       & 0.51                          & 0.29                       & 0.31                   \\
                                  & parameters & 7,174                         &  \hl{0.73}                      &  \hl{0.69}                          & 0.34                       & 0.34                          & 0.05                       & 0.00                          & 0.01                       & 0.00                   \\
                                  & responses  & 605                           &  \hl{0.99}                       &  \hl{0.97}                          & 0.70                       & 0.86                          & 0.89                       & 0.46                          & 0.27                       & 0.29                   \\

\hline
\end{tabular}

\label{tab:results-corrected-gt}

\end{center}

\vspace{-0.6cm}

\end{table*}

\subsection{Methods}
Only \ourapproach{} correctly identified all methods.
It generated two OpenAPI descriptions with the correct subsets of methods for the two Spring profiles in project \projectCwa{}.
Furthermore, \ourapproach{} correctly captured the regular expression constraints of parameters defined in the paths of project~\projectOcvn{} and removed them from the path string.

\ourapproach{} and Respector both obtained an overall precision and recall of 100\%. 
While Respector had a precision of 60\% for both profiles in \projectCwa{}, the project only contained three methods and, hence, did not have a noticeable impact on Respector's overall precision.
Prophet obtained 93\% overall precision and 51\% recall. Its results were only comparable to \ourapproach{} and Respector for two projects, \projectUrCodebin{} and \projectProxyPrint{}. Moreover, Prophet did not detect any methods of the project \projectCwa{} correctly because it could not resolve the constants in the endpoint paths. We observed the same shortcoming for Prophet in project \projectQuartz{}.

\springdoc{} achieved a low overall precision and recall (29\% and 31\%, respectively). This is because we could not find a valid run configuration for the three projects \projectOcvn{}, \projectProxyPrint{}, \projectQuartz{}, and no run configurations for the profiles \textit{external} and \textit{internal} of \projectCwa{}.
For the three projects, for which we could provide a run configuration, \springdoc{} obtained comparable precision and recall values but still lower than the values of \ourapproach{}. We could not observe any noticeable differences between missed and identified methods of \springdoc{}.

\subsection{Parameters}
\label{sec:quant:endpoint-parameters}

We flattened the named schemas of \ourapproach{}, also considering the inheritance hierarchy, during the quantitative evaluation to compare them to the individual parameters of the GT+ (cf. Section~\ref{sec:gt-corrections:parameters}). 
Similarly, we flattened the objects of Respector and \springdoc{} during the evaluation. 

\ourapproach{} correctly identified and converted the \texttt{@ModelAttribute} annotations in project \projectOcvn{}. Furthermore, it detected all parameters annotated with the \texttt{@RequestBody} annotation in the projects \projectCatwatch{} and \projectProxyPrint{}.
\ourapproach{} obtained an overall precision of 73\% and recall of 69\%, compared to Respector's 34\% for both, precision and recall. 
Prophet detected many parameters with simple types but could not translate parameter names based on the parameter annotation, resulting in low precision and recall. %
Similarly, \springdoc{} failed to correctly extract many parameters. 
\ourapproach{} missed several parameters in four projects due to the following reasons.

\subsubsection{No explicit parameter annotation}
Project \projectOcvn{} contains many parameters that do not contain an explicit parameter annotation, e.g., for paging. \ourapproach{} did not detect such parameters.

\subsubsection{HTTP servlet objects} %
Two projects use raw HTTP request and response objects encapsulating arbitrary parameters at runtime.
\projectOhsome{} exclusively uses \texttt{HttpServletRequest}, %
and \projectProxyPrint{} uses \texttt{WebRequest}.
\ourapproach{} could not detect the actual parameters encapsulated in such objects without fully analyzing the source code control and data flow.
Similarly, \springdoc{} ignored \texttt{HttpServletRequest} and \texttt{HttpServletResponse} objects per default~\cite{springdocOpenAPILibrary}.
Hence, \ourapproach{} and \springdoc{} did not report any parameters in project \projectOhsome{}, resulting in 0\% recall and precision. 
For project \projectProxyPrint{}, \ourapproach{} missed several parameters, and \springdoc{} could not generate the OpenAPI description.
Prophet naively reported all HTTP request and response objects as the parameters, resulting in 0\% precision and recall for project \projectOhsome{}.

\subsubsection{Json property annotations}
\ourapproach{} incorrectly identified the eight \texttt{@JsonProperty}-annotated parameters in project \projectUrCodebin{}, similar to Respector and Prophet. By ignoring the name translation, it created FP and FN findings.
However, these instances are negligible, considering the total number of parameters for all projects. 
We leave the handling of this annotation to future work.

\subsection{Responses} 
\ourapproach{} improved the detection of responses by correctly returning \textit{200 OK} for void methods and methods returning null and considering exception handling annotations.
It achieved an overall precision of 99\% and recall of 97\% for identifying responses, compared to Respector's 70\% precision and 86\% recall and Prophet's 89\% precision and 46\% recall. %
However, \ourapproach{} still missed responses in five projects.

\subsubsection{Propagated exceptions}
\ourapproach{} did not detect propagated exceptions from handler callees in the projects \projectCatwatch{} and \projectQuartz{}.
The \texttt{throws} statements in callees are outside of the analysis scope of \ourapproach{} and therefore not considered. 
Respector's symbolic execution identified many propagated exceptions but failed to translate them to the correct response codes.

\subsubsection{Propagated response codes}
Similar to propagated exceptions, \ourapproach{} did not detect \texttt{ResponseEntity} objects containing \textit{201 Created} that are returned from handler callees in the two profiles of project \projectCwa{}.

\subsubsection{Manual exception handling} 
In project \projectUrCodebin{}, \ourapproach{} and Respector reported \textit{500 Internal Server Error} instead of the manually translated response codes, simultaneously resulting in FP and FN findings, and hence, low precision and recall. 
The exception handlers return the response codes \textit{400 Bad Request}, \textit{404 Not Found}, and \textit{409 Conflict}.
However, we argue that these misclassifications are negligible because other projects use the \texttt{@ResponseStatus} annotation.
Prophet and \springdoc{} did not detect any error response codes for this project, and hence, created FN but no FP findings. This resulted in a comparable recall but higher precision than \ourapproach{} and Respector.

\subsection{Data model representation}
\newcommand{\labelResponseContentOurs}[0]{lst:responseContentOurs}
\newcommand{\labelResponseContentRespector}[0]{lst:responseContentRespector}

Finally, we evaluate the quality of the extracted data models.
\ourapproach{} described and referenced data model objects as named schemas in the \textit{\#/components/schemas} section. It used the \texttt{allOf} keyword to combine the schemas of derived objects and their superclasses while preserving the inheritance information. 
Listing~\ref{\labelResponseContentOurs} shows the schema for \texttt{InternalTestResult} of project \projectCwa{} generated by \ourapproach{}. 
The response references the named schema with the \texttt{\$ref} keyword in Line~6. This schema applies the \texttt{allOf} keyword in Line~10 and references its superclass in Line~11.
In comparison, Respector identified only a single field and described the object as an anonymous schema, as shown in Listing~\ref{\labelResponseContentRespector}. 
Prophet only reported the class name, and \springdoc{} could not analyze the project.

We discovered that only \ourapproach{} reported required fields of named schemas, as seen in Listing~\ref{\labelResponseContentOurs} in Line~20 because the corresponding fields are primitive Java types long (Line~25) and int (Line~32). The other approaches missed these constraints in a total of 24 named schemas.

\begin{figure}[htbp]
\input{evaluation/response-content-listing-ours}
\input{evaluation/response-content-listing-respector}
\vspace{-0.6cm}
\end{figure}

\noindent{\bf Answer RQ2}:
\ourapproach{} achieved the highest overall precision and recall when identifying methods, parameters, and responses. It outperformed the second-best approach, Respector, with an absolute 39\% higher precision and 35\% higher recall when identifying parameters and a 29\% higher precision and 11\% higher recall when identifying responses.
Moreover, \ourapproach{} is the only approach retaining inheritance information of data models and the only static analysis approach considering inherited fields of responses.

\section{Runtime evaluation}
\label{sec:eval-runtime}

In this section, we assess the runtime of \ourapproach{} and compare it to the other approaches to answer RQ3.
We do not consider \springdoc{} for this evaluation because it requires executing the service under analysis and exposes the OpenAPI description as an additional method. Hence, the runtime performance of \springdoc{} mostly depends on the service's runtime performance.

\subsection{Evaluation setup}
We executed all analyses on a virtual machine running Ubuntu 22.04 LTS, with four available cores of an AMD EPYC 7H12 CPU at 2.6 GHz and 16 GB of memory.
For each approach, we analyzed each project five times and report the average runtime per project.

\ourapproach{} parses the Java source code with Spoon, which is configured to cache project class paths in temporary files. This speeds up repeated analyses of the same project.
For this evaluation, we deleted all Spoon cache files before executing each analysis run, i.e., we measured the cold-start time of \ourapproach{}.
Respector is the only approach that requires bytecode for its static analysis. We compiled all projects before measuring Respector's analysis runtime, i.e., we excluded this preprocessing step from the measurements. %

Prophet is published as a Spring Boot web service. It receives the directory path(s) to one or multiple projects as an API call and returns the custom JSON format in the response. 
For this evaluation, we did not consider the one-time effort to start the Prophet web service and did not measure the conversion from the custom JSON format to the OpenAPI description. %
For each project, we measured the time taken to execute one \textit{curl} command and to write the response body to the terminal output. 

\subsection{Runtime performance}

We present the runtime results in Table~\ref{tab:runtime-performance}.
Prophet analyzed the projects the fastest, with an average runtime of 0.4 seconds and a median of 0.2 seconds. However, Prophet only analyzed handlers and did not generate any data model schemas.
\ourapproach{} obtained an average runtime of 11.0 seconds and a median of 7.6 seconds. It analyzed six projects in less than 10 seconds and one project, \projectOcvn{}, in 31.4 seconds. We discovered that project \projectOcvn{} is the only project with an invalid project configuration according to the IntelliJ Integrated Development Environment (IDE). Hence, we expect Spoon to take more time parsing the project correctly.
For the other projects, we observed that the runtime of \ourapproach{} increases approximately linearly with the project size in terms of Java LoC.
Notably, \ourapproach{} is the only approach without setup time. It is executable on the command line and only requires the Java source code as input.
Respector was the slowest approach, with an average runtime of 3,155.5 seconds, i.e., 52.6 minutes, and a median runtime of 447.6 seconds, i.e., 7.5 minutes. We attribute the extensive runtime to the symbolic execution analysis, which identifies parameter constraints that are not part of the OAS. %

\begin{table}[tbp]
\caption{Runtime results of the approaches in seconds for each project.}
\begin{center}
\begin{tabular}{|l|r|r|r|r|}

\hline
Project   & Java LoC & \ourapproach{} & Respector & Prophet \\
\hline

\projectCatwatch{}  & 6,454     & 7.6   & 551.9     & 0.2     \\
\projectCwa{}        & 3,616    & 6.8   & 6.3       & 0.1     \\
\projectOcvn{}      & 28,099    & 31.4  & 11,577.5  & 1.1     \\
\projectOhsome{}     & 10,597   & 9.8   & 9,499.3   & 0.7     \\
\projectProxyPrint{} & 6,052    & 8.8   & 447.6     & 0.3     \\
\projectQuartz{}     & 3,883    & 6.7   & 3.0       & 0.1     \\
\projectUrCodebin{} & 1,962     & 5.8   & 2.7       & 0.1     \\

\hline

Average             & 8,666.1   & 11.0  & 3,155.5   & 0.4     \\
Median              & 6,052     & 7.6   & 447.6     & 0.2     \\

\hline

\end{tabular}
\label{tab:runtime-performance}
\end{center}
\vspace{-0.6cm}
\end{table}

\noindent{\bf Answer RQ3}:
\ourapproach{} was slower than Prophet but significantly faster than Respector, which generates parameter constraints not part of the OAS. 
We argue that the runtime performance of \ourapproach{} is reasonable, especially considering its higher precision, recall, and data model quality compared to Prophet and Respector.

\section{Discussion}
\label{sec:discussion}
In this section, we discuss the implications and the internal and external threats to the validity of our evaluation.

\subsection{Implications}
\ourapproach{} offers higher applicability for practitioners than existing approaches for generating OpenAPI descriptions.
This is because Prophet and \springdoc{} require additional steps that complicate their usability.
In particular, Prophet was published as a web service whose analysis is initiated via an API call, and \springdoc{} requires running the service under analysis. Furthermore, Prophet outputs a custom format that cannot be directly used as an OpenAPI description. 
Respector simplified the generation of OpenAPI descriptions by only requiring bytecode. However, it suffers from a long runtime due to its symbolic execution, limiting its applicability.

In contrast, \ourapproach{} is implemented as a command line tool that achieves adequate runtimes while providing high precision, recall, and data model representation quality. In this regard, it significantly outperforms existing approaches. Furthermore, \ourapproach{} can be directly integrated into the development process by running it anytime during the implementation phase without needing the bytecode or a valid run configuration.
To initiate adoption by software developers, we also publish \ourapproach{} as GitHub action and GitLab template\footnote{Removed for anonymity}, enabling its use in continuous integration workflows.

\subsection{Threats to validity}
\label{sec:threats} 

To mitigate threats to \textbf{internal validity}, we used an existing dataset with seven Spring Boot web services curated by Huang et al.~\cite{10.1145/3597503.3639137} for evaluating the precision and recall of our generated OpenAPI descriptions. 
We discovered many mistakes in the dataset's original ground truth, and the first and second authors inspected the source code of the corresponding projects to discuss and improve the ground truth. 
We publish our improved ground truth, the web service projects, the \ourapproach{} source code, and the reported detection precision and recall metrics as a replication package~\cite{replication-package} for transparency.

To mitigate threats to \textbf{external validity}, we used seven Java Spring Boot projects 
that used multiple coding styles, third-party libraries, e.g., Project Lombok~\cite{projectlombokProjectLombok}, and request handling paradigms, e.g., servlet objects and annotation-based programming. %
Additionally, the projects comprised various numbers of endpoint paths, HTTP methods, parameters, and responses to be detected.
Accordingly, we demonstrated our approach's applicability with the Spring Boot framework. 
Our approach can be extended to other Java frameworks by adopting the analyzed annotations. For instance, the Jersey framework uses the \texttt{@GET} annotation instead of Spring's \texttt{@GetMapping} annotation~\cite{oracleDevelopingJax}. We provide the source code of our approach to allow other researchers to do that.

\section{Related work}
\label{sec:relatedwork}

In this section, we discuss related work on generating OpenAPI descriptions.
According to Lercher et al.~\cite{LERCHER2024112110}, OpenAPI is the de-facto standard in the industry to describe REST APIs.
Many existing approaches require executing the service to generate its OpenAPI description and, hence, require domain knowledge to create a valid run configuration of the service and the infrastructure resources to run it.

Yandrapally et al.~\cite{10.1109/ICSE48619.2023.00167} proposed ApiCarv, a test suite generation approach that executes an existing UI test suite for a web service in the browser, infers additional UI tests, and generates the OpenAPI description for all identified calls.
SwaggerHub Explore~\cite{swaggerAPIExploration}
and 
AppMap~\cite{appmapAppMap} %
are commercial products generating the OpenAPI description by recording REST API calls at runtime.
Serbout et al.~\cite{10.1007/978-3-031-36889-9_4} proposed ExpressO, an approach to generate OpenAPI descriptions for JavaScript's Express.js framework. They intercepted the initialization stage of the service at startup and stored the initialized methods and their handler source code.

SpringFox~\cite{githubGitHubSpringfoxspringfox} generated OpenAPI descriptions for projects developed with the Java Spring framework. 
However, the last contribution on GitHub was four years ago, and it was superseded by \springdoc{}~\cite{githubGitHubSpringdocspringdocopenapi}. 
Both tools use Java reflection to analyze Spring annotations at the time of service startup. Both require adding the corresponding dependency to the project and expose the OpenAPI description as an additional service method.
Swagger Core~\cite{githubGitHubSwaggerapiswaggercore} generates OpenAPI descriptions for web services developed with the Java Jersey framework. Again, it uses runtime reflection and exposes an additional method for accessing the OpenAPI description.

Recently, %
Huang et al.~\cite{10.1145/3597503.3639137} proposed Respector, the first approach to generate OpenAPI descriptions by statically analyzing Java bytecode.
They compared Respector to AppMap, Swagger Core, SpringFox, and \springdoc{} and outperformed all four tools for identifying methods, parameters, and responses. 
Additionally, they extracted parameter constraints by symbolically executing the bytecode. 
Furthermore, Cerny et al.~\cite{Cerny2024} proposed Prophet to visualize microservice architectures reconstructed from the source code. They report the REST APIs of services in an intermediate format, subsequently used to map API calls to the corresponding service methods. For comparison reasons, we converted the intermediate format to OpenAPI descriptions.

To the best of our knowledge, no approach exists that can generate an OpenAPI description accurately representing Spring profile configurations, exception handling, and data models, i.e., object and inheritance information. 
Furthermore, many existing approaches require injection into the running service to generate the OpenAPI description, which complicates the automated analysis. \ourapproach{} addresses these shortcomings and, as a result, generates more accurate and detailed OpenAPI descriptions.

\section{Conclusion}
\label{sec:conclusion}
In this paper, we proposed \ourapproach{}, our approach for automatically generating accurate OpenAPI descriptions for REST API web services from Java Spring Boot source code.
\ourapproach{} addresses current shortcomings of the state-of-the-art approaches Respector, Prophet, and \springdoc{}. 
It is the first approach that considers Spring's profiles and exception handling by generating one OpenAPI description per profile configuration and translating potential exceptions in handlers, i.e., Java methods, into HTTP response codes. 
Moreover, \ourapproach{} is the first approach that accurately represents parameter and response objects, i.e., the data models of the REST API. 

We evaluated \ourapproach{} on seven Java Spring Boot projects and compared the results to Respector, Prophet, and \springdoc{}. 
The results show that \ourapproach{} outperforms the second-best approach, Respector, with a 39\% higher precision and 35\% higher recall when identifying parameters and a 29\% higher precision and 11\% higher recall when identifying responses. 
Furthermore, it is the only static analysis approach that considers the inherited fields for all data models.
We observed that \ourapproach{} achieves these improvements at a median runtime of 7.6 seconds and maximum runtime of 31.4 seconds. 

Future work concerns handling the missed annotations, e.g., JsonProperty, and detecting propagated response codes and exceptions to further improve the precision and recall of our approach. Furthermore, we plan to extend our approach to support other Java frameworks, such as Jersey~\cite{eclipseee4jEclipseJersey} and Micronaut~\cite{Micronaut}.

\bibliographystyle{ieeetr}
\bibliography{sources}

\begin{thebibliography}{10}

\bibitem{fielding2000rest}
R.~T. Fielding, ``Rest: architectural styles and the design of network-based
  software architectures,'' {\em Doctoral dissertation, University of
  California}, 2000.

\bibitem{openapisOpenAPISpecification}
{OpenAPI Initiative}, ``{O}pen{A}{P}{I} {S}pecification v3.1.0.''
  \url{https://spec.openapis.org/oas/v3.1.0.html}.
\newblock [Accessed 10-10-2024].

\bibitem{LERCHER2024112110}
A.~Lercher, J.~Glock, C.~Macho, and M.~Pinzger, ``Microservice api evolution in
  practice: A study on strategies and challenges,'' {\em Journal of Systems and
  Software}, vol.~215, p.~112110, 2024.

\bibitem{openapiOpenAPIToolsOpen}
{APIs You Won't Hate}, ``{O}pen{A}{P}{I}.{T}ools.''
  \url{https://openapi.tools}.
\newblock [Accessed 10-10-2024].

\bibitem{10.1145/3184558.3188740}
I.~Koren and R.~Klamma, ``The exploitation of openapi documentation for the
  generation of web frontends,'' in {\em Companion Proceedings of the The Web
  Conference 2018}, WWW '18, (Republic and Canton of Geneva, CHE),
  p.~781–787, International World Wide Web Conferences Steering Committee,
  2018.

\bibitem{PENG20181313}
C.~Peng, P.~Goswami, and G.~Bai, ``Fuzzy matching of openapi described rest
  services,'' {\em Procedia Computer Science}, vol.~126, pp.~1313--1322, 2018.
\newblock Knowledge-Based and Intelligent Information \& Engineering Systems:
  Proceedings of the 22nd International Conference, KES-2018, Belgrade, Serbia.

\bibitem{10.1007/978-3-030-50578-3_40}
H.~Ed-Douibi, G.~Daniel, and J.~Cabot, ``Openapi bot: A chatbot to help you
  understand rest apis,'' in {\em Web Engineering} (M.~Bielikova, T.~Mikkonen,
  and C.~Pautasso, eds.), (Cham), pp.~538--542, Springer International
  Publishing, 2020.

\bibitem{9159071}
S.~Karlsson, A.~Čaušević, and D.~Sundmark, ``Quickrest: Property-based test
  generation of openapi-described restful apis,'' in {\em 2020 IEEE 13th
  International Conference on Software Testing, Validation and Verification
  (ICST)}, pp.~131--141, 2020.

\bibitem{10.1145/3293455}
A.~Arcuri, ``Restful api automated test case generation with evomaster,'' {\em
  ACM Trans. Softw. Eng. Methodol.}, vol.~28, Jan. 2019.

\bibitem{10.1145/3639476.3639769}
M.~Kim, T.~Stennett, D.~Shah, S.~Sinha, and A.~Orso, ``Leveraging large
  language models to improve rest api testing,'' in {\em Proceedings of the
  2024 ACM/IEEE 44th International Conference on Software Engineering: New
  Ideas and Emerging Results}, ICSE-NIER'24, (New York, NY, USA), p.~37–41,
  Association for Computing Machinery, 2024.

\bibitem{githubGitHubSpringdocspringdocopenapi}
springdoc, ``{springdoc-openapi}.''
  \url{https://github.com/springdoc/springdoc-openapi}.
\newblock [Accessed 10-10-2024].

\bibitem{githubGitHubSwaggerapiswaggercore}
{Swagger}, ``{Swagger Core}.''
  \url{https://github.com/swagger-api/swagger-core}.
\newblock [Accessed 10-10-2024].

\bibitem{10.1109/ICSE48619.2023.00167}
R.~Yandrapally, S.~Sinha, R.~Tzoref-Brill, and A.~Mesbah, ``Carving ui tests to
  generate api tests and api specification,'' in {\em Proceedings of the 45th
  International Conference on Software Engineering}, ICSE '23, p.~1971–1982,
  IEEE Press, 2023.

\bibitem{10.1007/978-3-031-36889-9_4}
S.~Serbout, A.~Romanelli, and C.~Pautasso, ``Expresso: From express.js
  implementation code to openapi interface descriptions,'' in {\em Software
  Architecture. ECSA 2022 Tracks and Workshops} (T.~Batista, T.~Bure{\v{s}},
  C.~Raibulet, and H.~Muccini, eds.), (Cham), pp.~29--44, Springer
  International Publishing, 2023.

\bibitem{10.1145/3597503.3639137}
R.~Huang, M.~Motwani, I.~Martinez, and A.~Orso, ``Generating rest api
  specifications through static analysis,'' in {\em Proceedings of the IEEE/ACM
  46th International Conference on Software Engineering}, ICSE '24, (New York,
  NY, USA), Association for Computing Machinery, 2024.

\bibitem{springSpringBoot}
{Spring}, ``{S}pring {B}oot.'' \url{https://spring.io/projects/spring-boot}.
\newblock [Accessed 10-10-2024].

\bibitem{eclipseee4jEclipseJersey}
{Eclipse Foundation}, ``{E}clipse {J}ersey.''
  \url{https://eclipse-ee4j.github.io/jersey/}.
\newblock [Accessed 10-10-2024].

\bibitem{appmapAppMap}
AppMap, ``{A}pp{M}ap.'' \url{https://appmap.io/}.
\newblock [Accessed 10-10-2024].

\bibitem{githubGitHubSpringfoxspringfox}
springfox, ``{SpringFox}.'' \url{https://github.com/springfox/springfox}.
\newblock [Accessed 10-10-2024].

\bibitem{Cerny2024}
T.~Cerny, A.~S. Abdelfattah, J.~Yero, and D.~Taibi, ``From static code analysis
  to visual models of microservice architecture,'' {\em Cluster Computing},
  vol.~27, pp.~4145--4170, Jul 2024.

\bibitem{springProfilesSpring}
{Spring}, ``{P}rofiles.''
  \url{https://docs.spring.io/spring-boot/reference/features/profiles.html}.
\newblock [Accessed 10-10-2024].

\bibitem{githubEMBjdk_8_mavencsrestoriginalcatwatchMaster}
{The Zalando Incubator}, ``{CatWatch}.''
  \url{https://github.com/WebFuzzing/EMB/tree/master/jdk_8_maven/cs/rest/original/catwatch}.
\newblock [Accessed 10-10-2024].

\bibitem{githubEMBjdk_11_mavenemembeddedrestcwaverificationMaster}
{Deutsche Telekom AG}, ``{Corona-Warn-App Verification Server}.''
  \url{https://github.com/WebFuzzing/EMB/tree/master/jdk_11_maven/em/embedded/rest/cwa-verification}.
\newblock [Accessed 10-10-2024].

\bibitem{githubEMBjdk_8_mavencsrestguiocvnMaster}
{Development Gateway}, ``{Open Contracting Vietnam (OCVN)}.''
  \url{https://github.com/WebFuzzing/EMB/tree/master/jdk_8_maven/cs/rest-gui/ocvn}.
\newblock [Accessed 10-10-2024].

\bibitem{githubGitHubGIScienceohsomeapi}
{GIScience Research Group and HeiGIT}, ``{Ohsome API}.''
  \url{https://github.com/GIScience/ohsome-api}.
\newblock [Accessed 10-10-2024].

\bibitem{githubEMBjdk_8_mavencsrestoriginalproxyprintMaster}
ProxyPrint, ``{proxyprint-kitchen}.''
  \url{https://github.com/WebFuzzing/EMB/tree/master/jdk_8_maven/cs/rest/original/proxyprint}.
\newblock [Accessed 10-10-2024].

\bibitem{githubGitHubFabioformosaquartzmanager}
{Fabio Formosa}, ``{Quartz Manager}.''
  \url{https://github.com/fabioformosa/quartz-manager}.
\newblock [Accessed 10-10-2024].

\bibitem{githubGitHubMathewEstafanousUrCodebinAPI}
{Mathew Estafanous}, ``{Ur-Codebin}.''
  \url{https://github.com/Mathew-Estafanous/Ur-Codebin-API}.
\newblock [Accessed 10-10-2024].

\bibitem{replication-package}
{Anonymous authors}, ``{Generating Accurate OpenAPI Descriptions from Java
  Source Code - Replication Package},'' Oct. 2024.
\newblock https://doi.org/10.5281/zenodo.13916835.

\bibitem{springModelAttributeSpring}
{Spring}, ``@{M}odel{A}ttribute.''
  \url{https://docs.spring.io/spring-framework/reference/web/webmvc/mvc-controller/ann-methods/modelattrib-method-args.html}.
\newblock [Accessed 10-10-2024].

\bibitem{openapisOpenAPISpecificationMapDict}
{OpenAPI Initiative}, ``{Model with Map/Dictionary Properties}.''
  \url{https://spec.openapis.org/oas/v3.1.0.html#model-with-map-dictionary-properties}.
\newblock [Accessed 10-10-2024].

\bibitem{pawlak:hal-01169705}
R.~Pawlak, M.~Monperrus, N.~Petitprez, C.~Noguera, and L.~Seinturier, ``{Spoon:
  A Library for Implementing Analyses and Transformations of Java Source
  Code},'' {\em {Software: Practice and Experience}}, vol.~46, pp.~1155--1179,
  2015.

\bibitem{openapisOpenAPISpecificationComponentsObject}
{OpenAPI Initiative}, ``{Components Object}.''
  \url{https://spec.openapis.org/oas/v3.1.0.html#components-object}.
\newblock [Accessed 10-10-2024].

\bibitem{openapisOpenAPISpecificationDataTypes}
{OpenAPI Initiative}, ``{Data Types}.''
  \url{https://spec.openapis.org/oas/v3.1.0.html#data-types}.
\newblock [Accessed 10-10-2024].

\bibitem{openapisOpenAPISpecificationPolymorphism}
{OpenAPI Initiative}, ``{Composition and Inheritance (Polymorphism)}.''
  \url{https://spec.openapis.org/oas/v3.1.0.html#composition-and-inheritance-polymorphism}.
\newblock [Accessed 10-10-2024].

\bibitem{springdocOpenAPILibrary}
{springdoc}, ``{What are the ignored types in the documentation?}.''
  \url{https://springdoc.org/#what-are-the-ignored-types-in-the-documentation}.
\newblock [Accessed 10-10-2024].

\bibitem{projectlombokProjectLombok}
{The Project Lombok Authors}, ``{P}roject {L}ombok.''
  \url{https://projectlombok.org/}.
\newblock [Accessed 10-10-2024].

\bibitem{oracleDevelopingJax}
{Oracle}, ``{Developing RESTful Web Services with Jersey}.''
  \url{https://docs.oracle.com/cd/E19776-01/820-4867/ggqnw/index.html}.
\newblock [Accessed 10-10-2024].

\bibitem{swaggerAPIExploration}
{Swagger}, ``{S}wagger{H}ub {E}xplore.''
  \url{https://swagger.io/tools/swaggerhub-explore/}.
\newblock [Accessed 10-10-2024].

\bibitem{Micronaut}
{Micronaut}, ``{Micronaut}.'' \url{https://micronaut.io/}.
\newblock [Accessed 10-10-2024].

\end{thebibliography}

\end{document}